\documentclass[a4paper,twoside]{article}

\usepackage{epsfig}
\usepackage{subcaption}
\usepackage{calc}
\usepackage{amssymb}
\usepackage{amstext}
\usepackage{amsmath}
\usepackage{amsthm}
\usepackage{multicol}
\usepackage{pslatex}
\usepackage{apalike}
\usepackage{algorithm2e}
\usepackage[bottom]{footmisc}
\usepackage{hyperref}
\usepackage{array}  
\usepackage{booktabs} 
\usepackage[table]{xcolor} 
\usepackage{ragged2e}
\usepackage[dvipsnames]{xcolor}
\usepackage{url}
\usepackage{tikz}
\usetikzlibrary{shapes, arrows, positioning}

\usepackage{SCITEPRESS}     

\newcommand{\gu}[1]{``#1''}	
\newcommand{\refig}[1]{Figure~\ref{fig:#1}}

\newcommand{\cds}{C\oe{}ur de Savoie\xspace}
\newcommand{\voltalis}{Voltalis\xspace}

\newcommand{\trvbase}{19.52}
\newcommand{\trvoffpeak}{16.35}
\newcommand{\trvpeak}{20.81}

\begin{document}

\title{Enhancing Collective Self-Consumption through Water Storage Heater Flexibility}

\author{\authorname{Pierre-Yves Massé\sup{1}, Maylis Duru\sup{2}, Benoit Couraud\sup{3}, Haicheng Ling\sup{1}\, Solal Bizeul\sup{2}, Hugo Roussel\sup{2}, Cléa Verdot\sup{4}, Mariane Vittoz\sup{5}, Estefania Alvarez\sup{6}\orcidAuthor{0009-0005-4470-5349}, Merlinda Andoni\sup{3}, Yann Rozier\sup{6},  Sonam Norbu\sup{3}, David Flynn\sup{3}, Erwin Franquet\sup{7}, Thibault Rihet\sup{1}}
\affiliation{\sup{1}Enogrid, 1 avenue Champ de Mars, CS 30019, 45074 Orléans CEDEX 2, France}
\affiliation{\sup{2}Voltalis, 75 avenue des Champs-Elysées, 75008 Paris, France}
\affiliation{\sup{3} James Watt School of Engineering, University of Glasgow, UK}
\affiliation{\sup{4}Communauté de communes \cds, Place Albert Serraz BP 40020, 73800 Montmelian cedex, France}
\affiliation{\sup{5}OPAC Savoie, 9 Rue Jean Girard Madoux, 73000 CHAMBERY, France}
\affiliation{\sup{6} Université Côte d’Azur, IMREDD, France}
\affiliation{\sup{7} Université Côte d’Azur, Polytech’Lab, France}
\email{
\{pierreyves.masse, haicheng, thibault\}@enogrid.com,
\{maylis.duru, solal.bizeul\}@voltalis.com, \{hugo.roussel26\}@gmail.com,
\{benoit.couraud, merlinda.andoni, sonam.norbu, david.flynn\}@glasgow.ac.uk,
\{ecoenergie\}@cc.coeurdesavoie.fr,
\{marianev16\}@gmail.com,
\{estefania.alvarez, yann.rozier, erwin.franquet\}@univ-cotedazur.fr,
} 
}


\abstract{While Renewable Energy Communities (RECs) and Collective Self-Consumption (CSC) schemes have emerged as promising tools to accelerate renewable energy adoption and support the net-zero transition, their full potential can only be realised when complemented by demand-side flexibility that aligns consumption with renewable generation. Water storage heaters can function as distributed thermal storage, absorbing excess renewable energy at the community level. 
This work quantifies both the benefits   of water storage heaters flexibility for energy consumers in a CSC community in France (such as energy bill reduction, increase of self-consumption), and the challenges related to the implementation and user acceptance. At the first stage, an annual simulation analysis is performed on a community of 41 households and a large solar PV plant, comparing a scenario without a CSC community, a scenario with a standard CSC community, and a scenario with a CSC community with flexibility from water storage heaters, which showed that an average benefit of 70€/year per household can be achieved due to flexibility and an increase of 6\% and 22\% of solar PV community self-consumption and self-production respectively. In the second stage, we present the results of the real-world deployment in the community, analysing its technical performance and user reception, and examine the main factors shaping user engagement and satisfaction.}

\onecolumn \maketitle \normalsize \setcounter{footnote}{0} \vfill

\section{\uppercase{Introduction}}
\label{sec:introduction}
The decarbonisation of energy systems relies on sustained investment in renewable electricity generation, particularly at the distributed and community scale. However, the significant reduction of export tariffs for small-scale renewable installations in many European countries, including France~\cite{cre2024pv}, has significantly weakened the economic attractiveness of distributed generation (DG) such as solar-PV. As a result, new business models are required to ensure the continued deployment of local renewable energy sources (RES).

In this context, Collective Self-Consumption (CSC) has emerged in France as a legally defined framework enabling groups of consumers and producers to locally share or trade electricity generation~\cite{codeenergieL315-2}. This regulatory scheme allows locally produced renewable electricity to be prioritised to supply local consumption within a defined geographic perimeter. To facilitate CSC operations, the French Distribution System Operator (DSO) provides dedicated support, notably by collecting metering data and by enabling the energy sharing rules requested by the community. 

Despite regulatory and operational enablers, CSC schemes often fail to fully absorb local RES production. Temporal mismatches between generation and demand frequently lead to surplus exports to the grid, which currently are characterised by limited financial rewards, thereby reducing the expected economic of DG. This limitation has motivated the evolution of CSC initiatives towards more active and flexible configurations, in which end-users adapt their consumption patterns to periods of local overproduction.

Among the available options for flexibility procurement, residential thermal loads, particularly electric water storage heaters (WSH), represent a promising resource due to their flexibility potential and widespread deployment in many countries~\cite{di2023flexibility,liu2015single}. In France alone, more than 50\% of residential consumers have electric WSH according to 2014 data~\cite{HPT_Annex46_France_2020}. 
WSH can offer a range of flexibility services, ranging from shifting their load to periods with lower electricity prices to providing grid operators with ancillary services and peak shaving. These systems exhibit several desired characteristics: their potential and technological maturity is already widely demonstrated for diverse use cases, especially for grid-wide applications, they represent a low-cost energy storage option, have a fast response time, and do not require large investments for retrofitting with communication and control systems (equipped systems do not exceed standard WSH costs by more than 10\%-20\%)~\cite{di2023flexibility}.  Activating such loads during periods of local renewable surplus can increase local self-consumption and enhance overall system efficiency and economic benefits.   

In the literature, the benefits from  Demand Response (DR) strategies have been simulated for Italian Renewable Energy Communities (RECs) in \cite{ercoli2025demand}, optimising space heating and cooling in buildings to increase photovoltaic (PV) self consumption and reduce electricity costs. However, real life implementation of such schemes faces significant obstacles, since real time control of space heating systems to track optimisation outputs is more complex than controlling water storage heaters (WSHs), which are thermally insulated and therefore more suitable for flexible operation. The potential of energy communities to deliver DSO services was also explored in \cite{askeland2025smart}, who showed a 13.1\% grid capacity need reduction. However, the integration of active demand-side flexibility within CSC schemes raises several open questions. \emph{First}, the actual benefits that energy community members can expect from their flexibility  remain to be quantified~\cite{allard2024quantifying}. For example, \cite{ercoli2025demand}  state that benefits depend on community composition, PV availability and energy sharing strategies. \emph{Second}, the feasibility of real-world deployment must be assessed, including technical constraints, operational risks, and coordination challenges. Similarly to barriers that DR faces in the general context, flexible CSC schemes suffer from lack of appropriate tools for system planners, complex communications~\cite{di2023flexibility}, market and regulatory fragmentation, difficulty on establishing baselines, DSO integration~\cite{le2023developing}, complex communications and privacy concerns~\cite{d2022exploiting}. \emph{Third}, the acceptability of flexibility-based control schemes by end-users is a critical challenge for long-term scalability and social sustainability. This social acceptance could be enhanced through transparent and fair benefit and energy allocation, but this needs to be demonstrated. Several research works have proposed fair energy allocation schemes~\cite{couraud2025fairnessenergydistributionmechanisms,madrigal2026improving}, however such mechanisms remain an open challenge that typically are not addressed in real deployments~\cite{askeland2025smart}. \cite{tomat2023insights} investigated the aspects that would enhance the flexibility effectiveness and social acceptability. Although more than 70\% of their survey respondents would join a DR scheme, driven mainly by the potential for cost savings, several groups reported resistance to changing habits. They also reported that acceptance improved when users were provided with plain language explanations. Similarly, \cite{luzzati2024energy} investigated the willingness of consumers to participate in flexibility and DR schemes using serious games, and compared the results of individual participation with group participation in a REC. They found that the sense of belonging to a wider group increased flexibility, and they attributed this result to peer influence, suggesting that combining flexibility with RECs can yield higher benefits.

To address these open challenges, this paper presents both a simulation-based theoretical analysis and early findings from a real-world experimental deployment, assessing the benefits and socio-technical challenges of enabling flexibility from residential WSH in a French CSC community of 41 members with one solar PV production owned by the municipality \cds. The study first evaluates the potential benefits of WSH flexibility through annual simulations based on real user consumption and production data from the community, focusing on maximising self-consumption at the community level. Building on these results, we report early findings from a real life implementation currently taking place at the community from \cds, where Enogrid, a leader in CSC deployments in France has provided software solutions for self-consumption optimisation, and Voltalis, a major stakeholder in residential demand side flexibility in France has provided flexibility activation devices  to control  WSH of several households within the community. The first phase of the experiment started in fall 2025, and lasted for 45 days. 
~We report the early experimental results from this first phase, which we analyse semi-quantitatively, and highlight key operational trends that point to the main challenges associated with larger scale deployment.
In particular, the paper incorporates feedback from participating end users to assess the acceptability of flexibility-enabled CSC schemes and to identify the main factors influencing user engagement and satisfaction.

The remainder of the paper is structured as follows. Section 2 presents the results of the annual simulation of flexibility activation in the French energy community. Section 3 reports on the real-world deployment of this flexibility at selected members’ premises, including the design of the required forecasting tools, while Section 4 examines the implementation results and discusses the acceptability of community-based demand side flexibility schemes.

\section{Benefit Quantification of WSH Flexibility in a French Energy Community}

In this section, we analyse the benefits expected in a real French energy community from implementing WSH flexibility combined with a CSC scheme. The community considered in this study comprises a single solar PV installation owned by the municipality of \cds in France, with a nominal capacity of 36 kWp. It serves 41 residential consumers located in the vicinity of the PV system, each equipped with a similar water storage heater (WSH).

\begin{figure}[t]
    \centering
    \includegraphics[width=\linewidth]{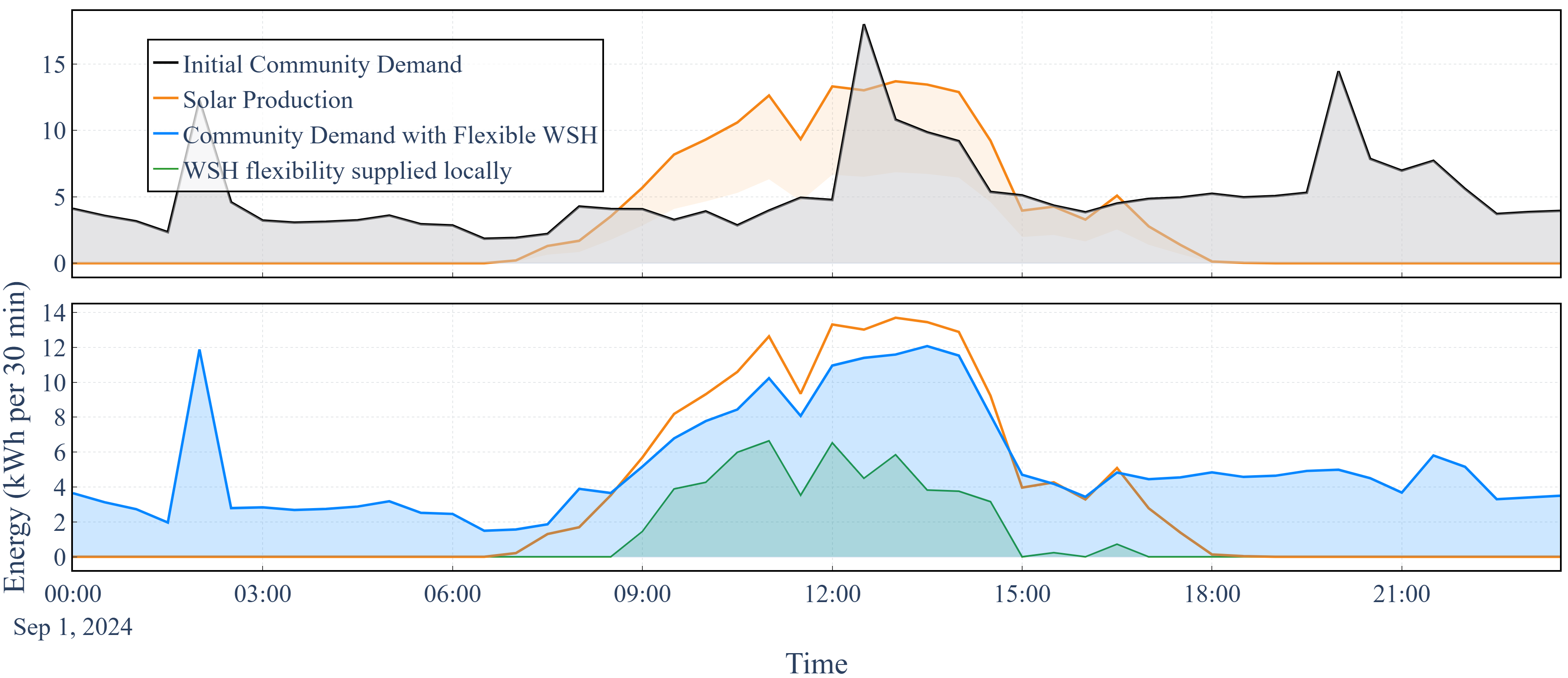}
    \caption{Daily load and production profiles of the community on a typical day, without flexibility (top) and with simulated flexibility (bottom).}
    \label{fig:daily_load_profile_flex}
\end{figure}

As defined in the CSC scheme, when local production is available, these consumers self consume the electricity generated by the community PV system at no cost, apart from network charges and taxes. Therefore, this configuration is assimilated to a collectively owned asset with shared benefits.  For electricity consumed during periods without solar PV production, consumers import electricity from the main grid. Two electricity supply tariff structures are observed within the community: 
\begin{itemize}
\item Fixed-price contracts, under which consumers pay \trvbase~c€/kWh for every kilowatt-hour imported from the grid.
\item  Time-of-Use (ToU) contracts (Peak–off-peak), under which consumers pay \trvoffpeak~c€/kWh during the following periods: 02:00-06:00, 12:30-14:30, and 20:00-22:00 and \trvpeak~c€/kWh during all other periods.
\end{itemize}
Sixty one percent of energy community members are currently subscribed to a time of use tariff. Flexible assets such as water storage heaters are automatically triggered during off peak periods. As shown in the upper part of Fig.~\ref{fig:daily_load_profile_flex}, which illustrates a typical daily load (in gray) and production profile of the energy community, three consumption peaks are observed at 2 AM, 12:30 PM, and 8 PM, due to the triggering of WSH. The figure also shows that PV production is not fully self consumed, mainly because of insufficient demand during periods of high solar generation. For example, although a consumption peak occurs around 12:30 PM, a significant share of PV production is still exported to the main grid before and after that time period.

Three steps were followed to quantify the potential benefits of aligning WSH with local PV production.
~First, we evaluate the baseline bills for each household as if there was no energy community (including the revenues from the production installation with an export tariff with a supplier). Then, we compute the benefits of aggregating these households into an energy community with a standard CSC scheme without demand-side flexibility, to assess the economic benefits from local self-consumption. Finally, we assess the additional impact of enabling flexibility within the CSC scheme.

\subsection{Baseline Scenario}
In the baseline scenario, all community solar PV production is sold to a supplier at an advantageous export tariff of 13 c€/kWh. This tariff is used for replication purposes in other countries but is no longer applicable in France, where export tariffs have been reduced to 8 c€/kWh at best for most installations in recent years, thereby strengthening the case for CSC based business models~\cite{cre2024pv}. Community members meet their electricity demand by importing energy from the main grid under their respective tariff structures: fixed price tariffs for 39\% of consumers and ToU tariffs for the remainder. Based on one year of monitored individual consumption and production data, and assuming an investment cost of 1.2 k€/kW for the 36 kWp PV system, the resulting simple payback period of the solar installation is estimated at 8.5 years. Similarly, the aggregated baseline electricity bill for the entire community amounts to 56 k€.
Results comparison with other scenarios are available in Table \ref{tab:scenario_comparison}, where the annual bill reduction for the baseline is computed as the ratio between the financial benefits from solar PV exports and the number of community members.

\subsection{Standard CSC Scenario}
In the standard CSC scenario, community members are allowed to self-consume the local solar PV production whenever their demand coincides with production, at no cost, apart from taxes and network charges for this local energy. Surplus electricity that cannot be consumed locally is exported to the grid and remunerated at the same 13 c€/kWh export tariff, as in the baseline scenario.

\begin{figure}[t]
    \centering
    \includegraphics[width=\linewidth]{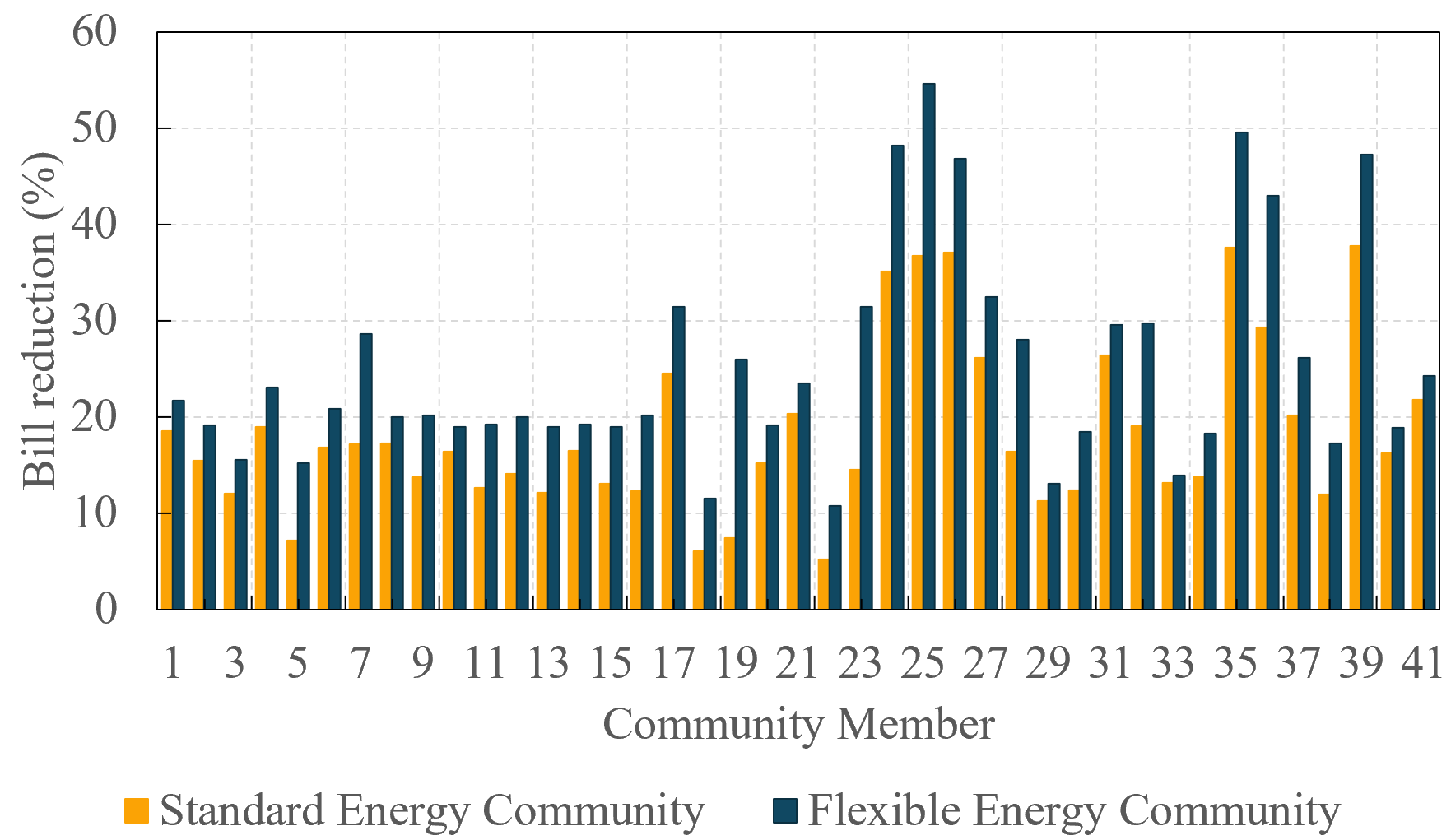}
    \caption{Comparison of individual bill reduction with respect to the baseline scenario (no community) in the two scenarios: community without flexibility (orange), and community with flexibility (blue). For instance, member 17 reduces their bill by 25\%  without flexibility, and by more than 30\% with flexibility.}
    \label{fig:individual_benefits_CSC}
\end{figure}

The allocation of locally produced electricity among community members follows the default allocation mechanism proposed by the DSO, namely the pro rata rule \cite{couraud2025fairnessenergydistributionmechanisms}. For each 30 minute time step, local energy is allocated to each consumer proportionally to their share of total community consumption during that time step. As a result, all community members experience reductions in their electricity bills that are proportional to the amount of energy they consume during production periods, as illustrated in orange in Figure~\ref{fig:individual_benefits_CSC}.
In the configuration with a collectively-owned asset, these savings can contribute to reimbursing the PV investment, reducing the simple payback period to 4.1 years. Overall, the CSC framework provides a stronger economic incentive for local renewable generation investment than the baseline case, even with an advantageous export tariff of 13c€/kWh.
~This is driven by the higher retail electricity price relative to the export tariff. 

\begin{table}[ht]
    \vspace{-1mm}
    \caption{Scenarios Comparison.}
    \label{tab:scenario_comparison}
    \small
    \centering
    \begin{tabular}{>{\raggedright\arraybackslash}m{1.5cm} 
                    >{\raggedright\arraybackslash}m{1.3cm} 
                    >{\raggedright\arraybackslash}m{1.3cm} 
                    >{\raggedright\arraybackslash}m{1.3cm}}
        \toprule[\heavyrulewidth]\toprule[\heavyrulewidth]
        \textbf{} & \textbf{Baseline} & \textbf{Standard CSC} & \textbf{Flexible CSC} \\
        \midrule
        \arrayrulecolor[gray]{0.8}

        Initial PV Investment & 41k€ & 41k€ & 41k€ \\
        \specialrule{0.01pt}{1pt}{1pt}

        Annual Production & 4.6MWh & 4.6MWh & 4.6MWh \\
        \specialrule{0.01pt}{1pt}{1pt}

        Annual Consumption & 17.2MWh & 17.2MWh & 17.2MWh \\
        \specialrule{0.01pt}{1pt}{1pt}

        Self-Consumption & 0\% & 19\% & 24.9\% \\
        \specialrule{0.01pt}{1pt}{1pt}

        Self-Production & 0\% & 71.3\% & 93.3\% \\
        \specialrule{0.01pt}{1pt}{1pt}

        Annual Bill Reduction per Household & 123€ & 190€ & 260€ \\

        \arrayrulecolor{black}
        \bottomrule[\heavyrulewidth]
    \end{tabular}
    \vspace{-3mm}
\end{table}

\begin{figure}[t]
    \centering
    \includegraphics[width=\linewidth]{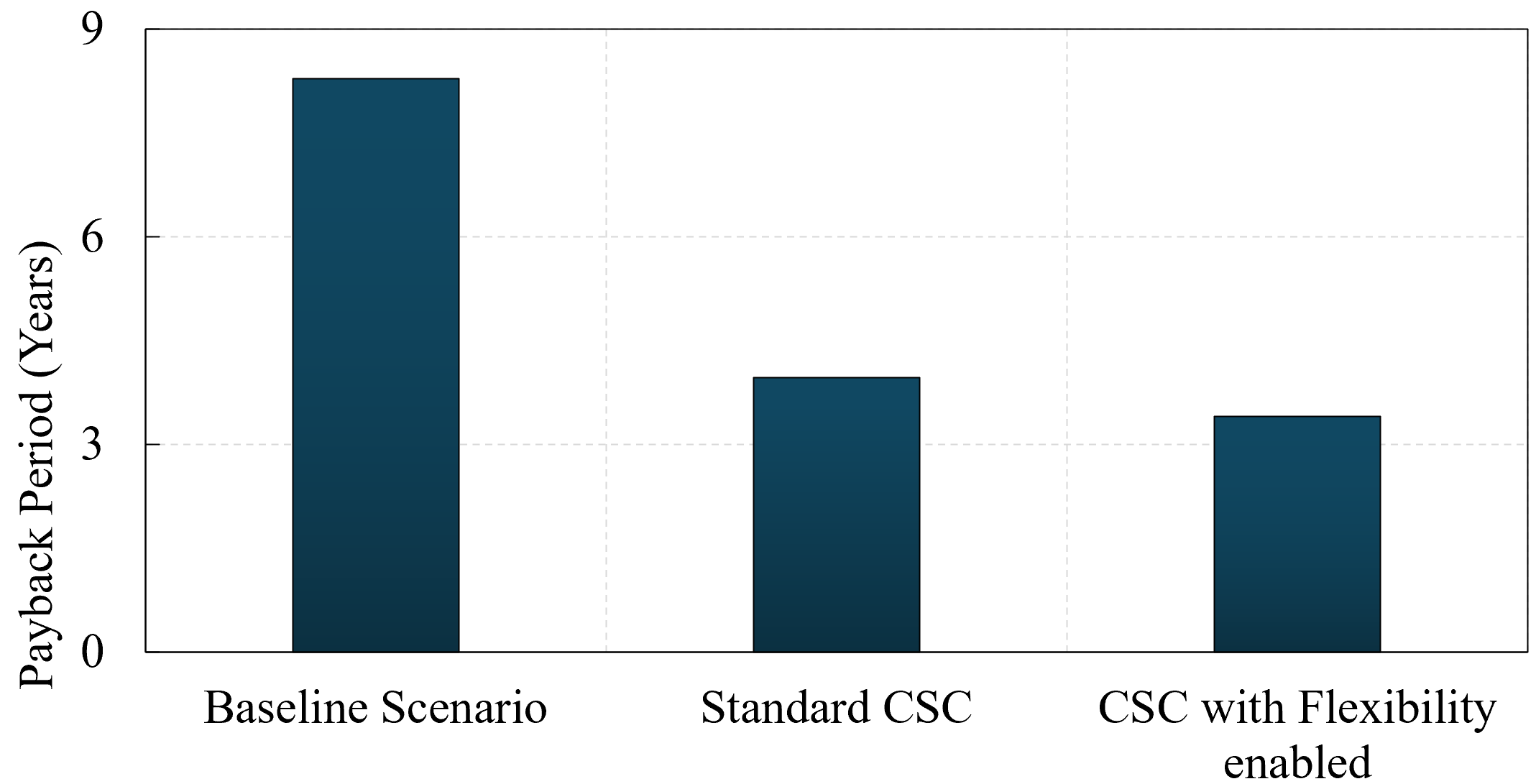}
    \caption{Comparison of payback periods among the scenarios considered.}
    \label{fig:payback_period}
\end{figure}

\subsection{CSC with Flexibility from WSH}
\label{sec:simulation_flex}
In the final scenario, we evaluate the benefits of enabling flexibility from residential WSH to increase community PV self-consumption, i.e. to reduce the PV surplus exported to the grid by shifting WSH consumption to periods of excess solar generation.

In this scenario analysis, an optimal flexibility case is considered, assuming real-time monitoring, perfectly accurate forecasting and seamless control of WSH loads, which provides a `best case scenario' and theoretical benchmark for quantifaction of the flexibility potential. In this idealised setup, the following overarching assumption is considered: all flexible WSH consumption can be shifted to time steps with available PV production, except for a guaranteed activation period for ToU consumers. For households with ToU tariffs, WSH activation is maintained during the first half hour of the night off-peak period at 2 AM to ensure sufficient hot water availability, particularly to cover morning hot water demand of working occupants, as identified by the weekdays activity analysis from~\cite{Bejannin_2020}. The remaining daily WSH energy demand is then reallocated to periods where PV production exceeds the community non-flexible load.

\begin{figure*}[t]
    \centering
    \includegraphics[width=0.85\textwidth]{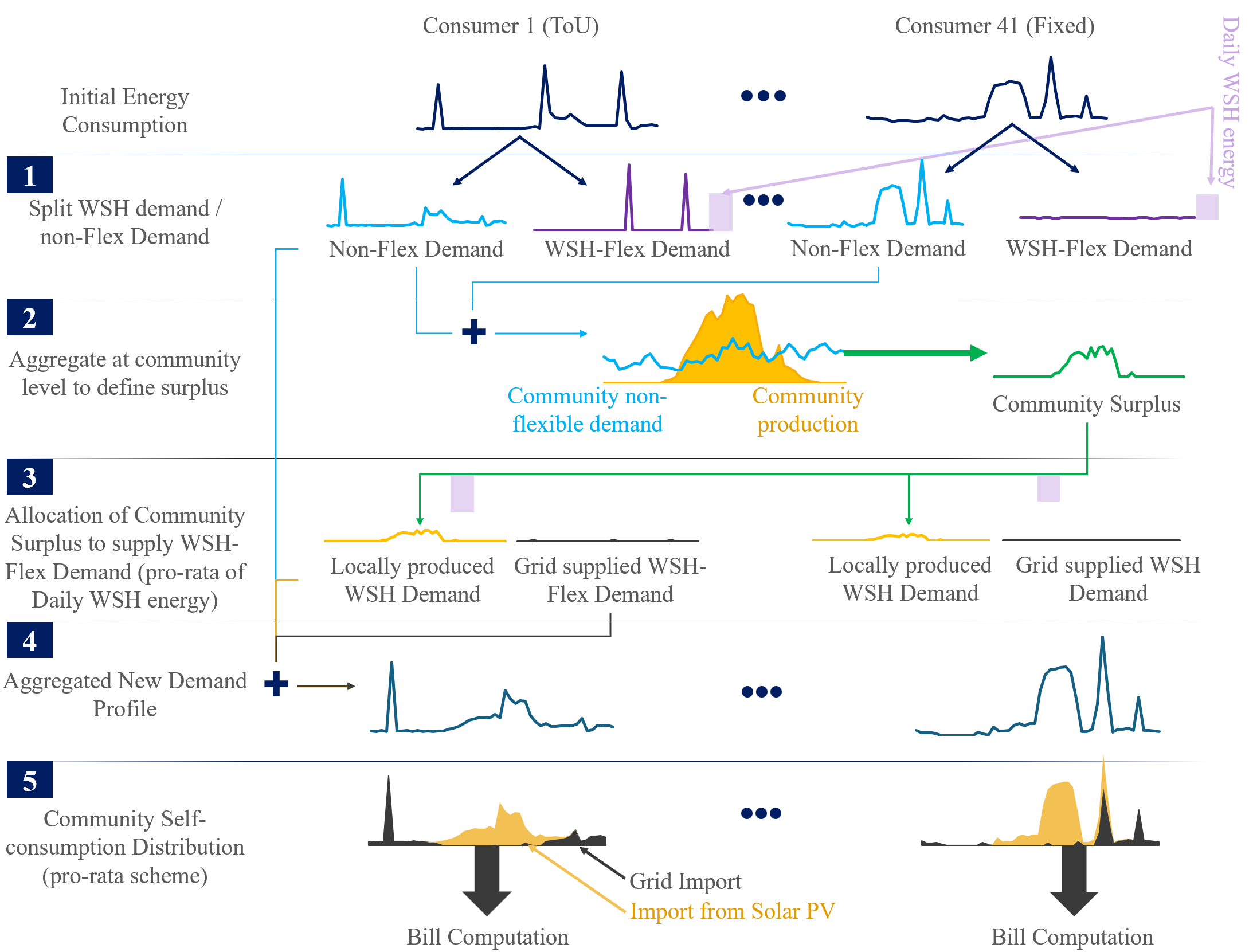}
    \caption{Process for WSH flexibility integration in the community.}
    \label{fig:simulation_process}
\end{figure*}

In more detail, the scenario analysis followed the procedure described below, and displayed in Fig. \ref{fig:simulation_process}:

\begin{enumerate}
\item \textbf{Split WSH flexible demand and non-flexible demand}: we aim to shift flexible WSH demand to times when solar production exceeds the community non-flexible demand. As a result, we need to estimate the flexible WSH demand from the total household consumption data available from smart meter monitoring, as no sub-metering was available for individual WSH load monitoring. 
~For consumers with ToU tariffs,  WSH power was identified through peak detection. A peak was defined as a load increase at the beginning of an off-peak period, greater than 0.8 kW relative to a baseline, computed as the average demand over the three time steps preceding the off-peak period. The WSH event was considered to last until demand returned below the baseline plus 0.3 kW. The daily WSH energy was computed as the sum of these detected peaks on that day. For consumers on fixed tariffs, for whom no systematic off-peak signal exists, daily WSH energy was approximated as the average WSH energy observed among ToU households, and assumed to be uniformly distributed within the day, reflecting the quasi periodic operation of thermostatically controlled heaters, as shown by the purple curves in the first step of Fig. \ref{fig:simulation_process}. Note that since these users face a constant tariff, this approximation does not affect baseline bill calculations.

A non-flexible baseline load was then constructed for each household (in light blue in Fig. \ref{fig:simulation_process}) by subtracting the estimated WSH profile from the measured load. For ToU households, the extracted peak profile was removed, apart from the peak at 2 AM. 

\item \textbf{Community Solar PV excess generation computation}: from individual  non-flexible demand profiles derived in Step 1, we computed the community aggregated non-flexible load by aggregating all households non-flexible demand curves. The PV community surplus was defined as the difference between PV production and the aggregated community non-flexible load. 

\item \textbf{Surplus allocation among consumers}: The available community surplus at each time step was then allocated to households according to a pro rata rule based on their daily flexible WSH energy requirement (in light purple in Fig. \ref{fig:simulation_process}). Households with higher daily WSH needs receive a larger share of the available PV surplus. This results in a new WSH load profile for each consumer (in orange), following this local energy distribution mechanism. Remaining WSH energy that cannot be supplied by the community surplus is then distributed between the off-peak time-slots following the household's WSH original demand profile for ToU households, as shown in the black curve in Fig. \ref{fig:simulation_process} or uniformly throughout the day for households with a fixed price tariff.

\item \textbf{New demand profile}: This new WSH load profile is then aggregated with the non-flexible demand profile to generate the new demand profile of the household, which includes the flexibility that will leverage the community surplus. 
\item \textbf{Economic assessment}: Finally, using these new load profiles, we  compute the energy bills for each consumer, where demand served by a share of the community PV production incurs no energy cost, while the remainder is charged with the corresponding tariff price.
\end{enumerate}

For one typical day, the resulting community demand pattern is shown at the bottom of Figure~\ref{fig:daily_load_profile_flex}, in blue.
~We can observe that the 2 AM peak load is maintained for households under ToU tariffs, while the remaining WSH load is shifted to periods of community PV surplus. As a consequence, midday excess production is absorbed and the previously observed 12:30 PM peak that exceeded PV generation disappears, increasing the self-production rate. The community load profile closely follows the solar PV production curve, with the blue curve nearly overlapping the orange curve in Figure~\ref{fig:daily_load_profile_flex}, indicating near optimal self-consumption.

Economically, enabling WSH flexibility further reduces the annual bill
~compared to standard CSC without flexibility. Several low-consumption households achieve up to 54\% annual bill reduction, although these correspond to dwellings without electric heating. On average, bill savings increase from 190 €/year/household under standard CSC to 260 €/year with flexibility. At the community level, this additional value further reduces the PV simple payback period from 4.1 years to 3.5 years, as shown in Fig. \ref{fig:payback_period}. In addition to economic gains, midday and evening load peaks are mitigated, contributing to improved local grid conditions.

While the results of the theoretical analysis look very promising, several limitations must be emphasised. First, the costs associated with enabling WSH flexibility, including installation and operation of control devices, were not included in the economic assessment, and should be compared with the additional gain relative to standard CSC of approximately 70€/year per household. Second, the scenario assumes perfect forecasting, full controllability, and flawless communication. It does not account for uncertainties, behavioral variability, control constraints, or technical failures that arise in real deployments, which will be discussed in Sections \ref{sec:experiment} and \ref{sec:sociotechnicalresults}. The next section compares these results with industry offerings.

\subsection{Comparison with Industry Alternatives for Flexibility}

Beyond the scenarios explored in the previous section, we compare our estimation of the gains through CSC water heater flexibility with two industry alternatives, one for producers, and one for consumers.
First, a producer may sell energy in flexibility schemes on wholesale and ancillary service markets. According to data from a French aggregator (market operator able to trade energy) \cite{elmy_flexibilite_enr_2024}, and assuming that PV panels of the CSC could have been used to participate in these markets,
~the average gains of a PV asset of a similar size as in our study, in the year 2024-2025, would have been 250€ or 6€/household/year.
~Second, the French branch of the electricity supplier Octopus have just launched a new flexibility offer on WSH for households: they offer to control WSH remotely, and make them heat during favorable periods. In exchange, they will refund 45€/household/year for fixed prices households, and 27€/household/year for households with ToU prices \cite{octopus_ballons_2026}. This is to be compared with the bills savings of 260€/year/household obtained from participation in a CSC energy community with flexibility, with an initial investment of $\sim$1k€ per member for solar PV installation. 
In both cases, we argue that it is more beneficial in the long term to invest in solar production and leverage the business case from CSC schemes. While these gains remain theoretical at this stage, community-based flexibility is an emerging topic, whose effectiveness is expected to improve in the future. As a result, this study together with the first phase of the experiment discussed in this work constitute important first steps in this promising direction.
The following section presents the setup of a real-world implementation of demand side flexibility in the studied community. 

\section{\uppercase{Real-World Pilot Setup}}
\label{sec:experiment}
This section presents the operational configuration and setup of the real-world implementation in the energy community from \cds.

\subsection{Recruitment of volunteers}
Volunteers for the experiment on WSH flexibility  were recruited  among the members of the community through a targeted mailing and phone campaign. The principles of the experiment were explained, and members were invited to express their interest through an online form.
~Overall, eight households, accounting for 25\% of the community, participated in the experiment. 

\subsection{Operational setup:}
\textbf{Smart control of electric devices.} The experiment was conducted through the installation of the standard \voltalis solution, which monitors and controls WSH either directly through the \voltalis control algorithm or via a dedicated application accessible to users. Users retain full control over their appliance and can override the control commands issued by \voltalis at any time through the application. 

\begin{figure}[t]
    \centering
    \includegraphics[width=\linewidth]{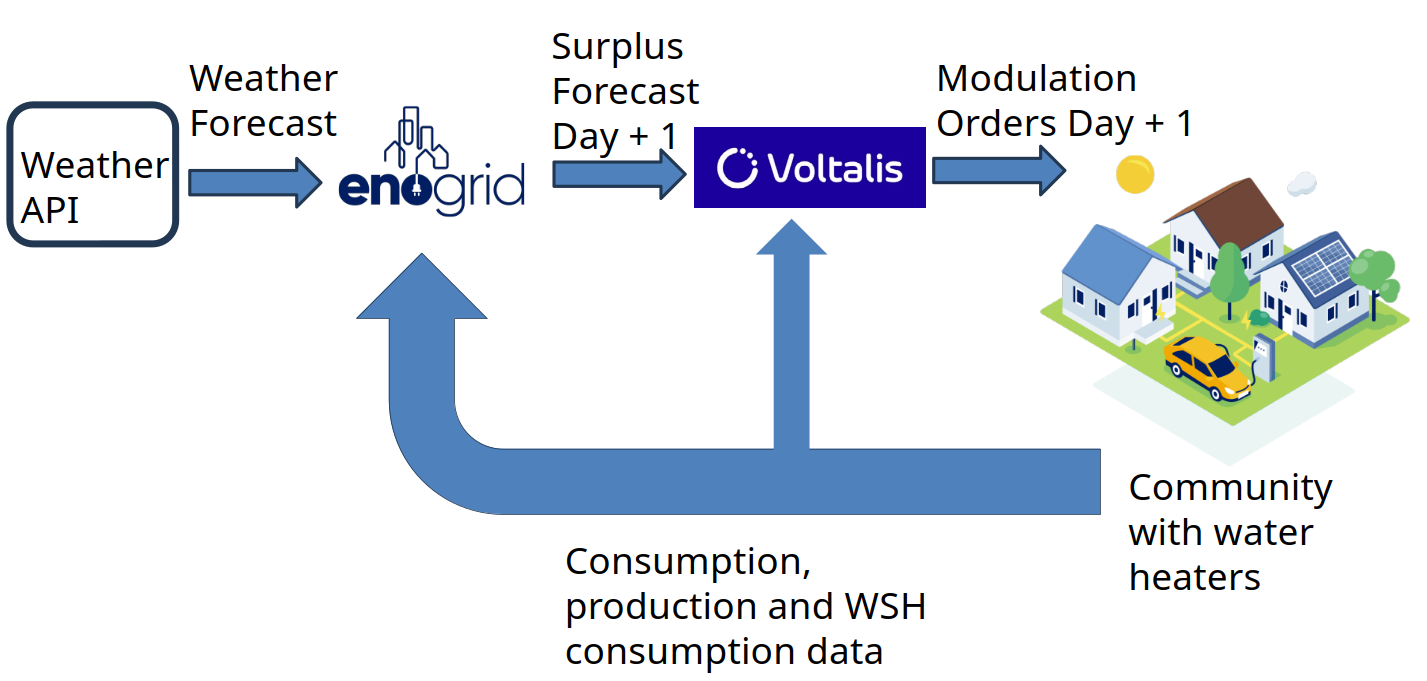}
    \caption{Experimentation setup}
    \label{fig:setup}
\end{figure}

Following the approach detailed in Section \ref{sec:simulation_flex}, the WSH of participating households were preferentially activated during periods of forecasted local solar surplus. Indeed, because there is no real-time monitoring of the households consumption (only of the WSH of flexible participants), the decision to activate the WSH relies exclusively on the day-ahead community surplus forecast, computed by Enogrid's forecasting tools as the difference between the day-ahead solar PV generation and community consumption forecasts. 

On a day ahead basis, when solar PV surpluses were forecast, a modulation plan was sent by Voltalis to the WSH for the day (see Fig. \ref{fig:setup}), whereby each 5 minutes period was labeled as either "modulated", in which case the WSH could not activate, or ``not modulated", in which case the WSH could activate at will. In addition, as described in Section \ref{sec:simulation_flex}, the WSH of households under ToU tariffs were systematically activated at the beginning of the 2 AM off-peak period to guarantee hot water availability in the morning.

Therefore, forecasting accuracy is critical to ensure that WSH flexibility is not activated during periods without an actual solar PV surplus. The next section details the forecasting approach used in the experiment. 

\subsection{Forecasting Surplus:}

Random Forest (RF) models were used to predict the day-ahead of consumption and PV production. Model accuracy is then evaluated using daily normalised error metrics to capture scale-independent performance.

\subsubsection{Random Forest Based Models and Input Data}
To support the scheduling of flexible loads, short-to medium-term forecasts of electricity consumption and PV production are computed using RF regression models. RF models are well suited to the CSC context, as they capture nonlinear relationships while remaining robust to noise, and are well adapted to medium-sized datasets. In particular, residential electricity demand is partly shaped by regulated peak–off-peak tariff structures, leading to relatively regular consumption patterns for flexible loads such as WSH, which represent a significant share of the household electricity demand. RF is an ensemble regression method that constructs multiple decision trees on random subsets of data and features, and averages their outputs to produce robust predictions. This ensemble structure enables RF to model complex nonlinear relationships and interactions between input variables effectively, while reducing overfitting compared to single decision trees \cite{Breiman2001}. In this work, key RF hyperparameters such as the number of trees and maximum tree depth are selected via cross-validation to balance bias and variance.

Two separate predictors are developed: one for household electricity consumption and one for PV production. Both models rely on historical energy data (production and consumption) and meteorological variables (ambient temperature, global solar irradiance, relative humidity, and cloud cover), complemented by lagging features to capture temporal dependencies. For the consumption predictor, additional temporal features are included to reflect human-driven usage patterns (hour of day, day of week, day of year, weekends and public holidays). For the production predictor, temporal features are limited to the hour of the day and the month of the year, as PV generation is mainly governed by daily cycles and seasonal patterns.
The consumption and production forecasts are combined to estimate the expected surplus.

\subsubsection{Forecast Accuracy}

Forecasting performance is evaluated by comparing predicted values with measured data over the evaluation period using normalised error metrics computed on a daily basis for both electricity consumption and PV production. In contrast to scale-dependent indicators, normalised metrics enable a meaningful comparison across variables with different magnitudes and physical characteristics.

Specifically, we employ the normalised Mean Absolute Error (nMAE) and the normalized Root Mean Square Error (nRMSE) \cite{solar5040048,Demir2025}, defined as:

\[
\mathrm{nMAE}
= \frac{1}{P_{\mathrm{nom}}}
\cdot
\frac{1}{N}
\sum_{t=1}^{N}
\left| \hat{P}_t - P_t \right|
\]

\[
\mathrm{nRMSE}
= \frac{1}{P_{\mathrm{nom}}}
\cdot
\sqrt{
\frac{1}{N}
\sum_{t=1}^{N}
\left( \hat{P}_t - P_t \right)^2
},
\]
where $P_{\mathrm{nom}}$ denotes a normalisation constant corresponding to the annual maximum observed value for each variable (annual peak consumption for electricity demand and installed peak capacity for PV production), $N$ is the number of time steps, $P$ is the ground truth (consumption or production), and $\hat{P}$ the corresponding forecast. The normalisation ensures scale independence while preserving physical interpretability.

\begin{figure}[t]
    \centering
    \includegraphics[width=\linewidth]{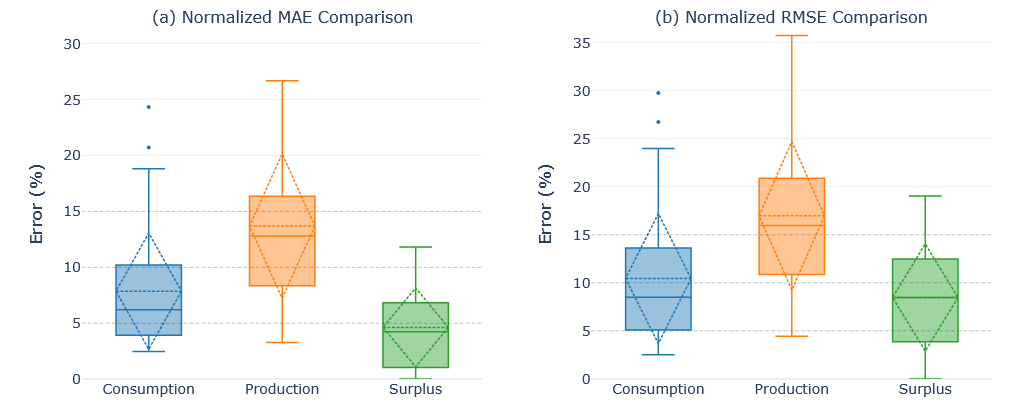}
    \caption{Distribution of daily forecasting accuracy metrics for electricity consumption (blue) and photovoltaic (PV) production (orange) and resulting surplus (green) over October 2025.}
    \label{fig:forecast_accuracy}
\end{figure}

\refig{forecast_accuracy} presents the distribution of daily normalised forecasting errors over October 2025 for consumption, production and surplus. Data are restricted to daytime, as this is the period of interest. Results are shown as boxplots in Fig.~\ref{fig:forecast_accuracy}. Extreme values are displayed as individual markers.

For electricity consumption, the median daily nMAE remains below 10\%, with an interquartile range indicating relatively stable performance across the evaluation period. The corresponding nRMSE values are moderately higher, reflecting occasional peak-load deviations that increase squared error contributions. A limited number of days exhibit significantly larger errors, due to irregular consumption usages. Overall, forecast performance is satisfactory for consumption. Errors are higher for PV production, with median nRSME above 15\%. This reflects in particular the sensitivity of production to sudden changes in weather, whereby production decreases due to clouds, which were not forecast accurately in advance. However, surplus forecast are better: errors in consumption and production may sometimes compensate, and moreover, whenever forecast production is lower than consumption, any further deterioration of production does not lead to worsening surplus forecast, as it already is zero. Overall, performance was sufficiently accurate to enable the predictive control of the WSH, but errors downgrade the performance of the control. Nevertheless, there remains scope for further improvement to enhance the accuracy of forecasts, and, consequently, the overall performance of the WSH flexibility system.

\section{Socio-technical Results}
\label{sec:sociotechnicalresults}
In this section, we present the results of the real world implementation described in the previous section and report the feedback from a post pilot user survey assessing the acceptability of the proposed solutions.

\subsection{Experimental Results}
In this section, we present the early findings of the first phase of the real-world deployment, which took place in Autumn 2025. 

\textbf{Unfolding of the experiment.} Phase A of the experiment ran from mid-September, until the end of October. WSHs were included in the experiment as soon as they were equipped with the Voltalis kit.
~Note that the experiment was interrupted for a fortnight at the beginning of October, due to operational issues, but resumed afterwards. These days where no control orders were sent were used as reference days to evaluate the influence of the control system.

\textbf{Key metrics.} To monitor the efficiency of the proposed solution, we monitored the following key indicators: At the community operation level, we monitor the self consumption rate, defined as the proportion of locally generated energy consumed by community members. However, due to the short duration of the experiment, the absence of a comparable baseline period, and the limited number of participants, it was not possible to robustly quantify the impact on self consumption. Then, at the household level we monitor the self-sufficiency rate, or the ratio of the consumption of the household which is covered by local production.

\textbf{Restrictions on ToU WSH.} After careful consideration, it was decided to prevent WSH demand shifts from off-peak to peak hours in the case of households under ToU tariffs, to negate financial risks that might have occurred due to the uncertainty of forecasts. Indeed, in the case of an erroneous forecast predicting an energy surplus at 11 AM when no surplus actually occurred, WSHs could have been activated during periods of peak prices without being supplied by local production. This would have exposed households to high electricity prices for the WSH load.
This restriction did not apply for consumers with fixed energy supplier rates, as this would not have financial impact.

\begin{figure}[t]
    \centering
    \includegraphics[width=\linewidth]{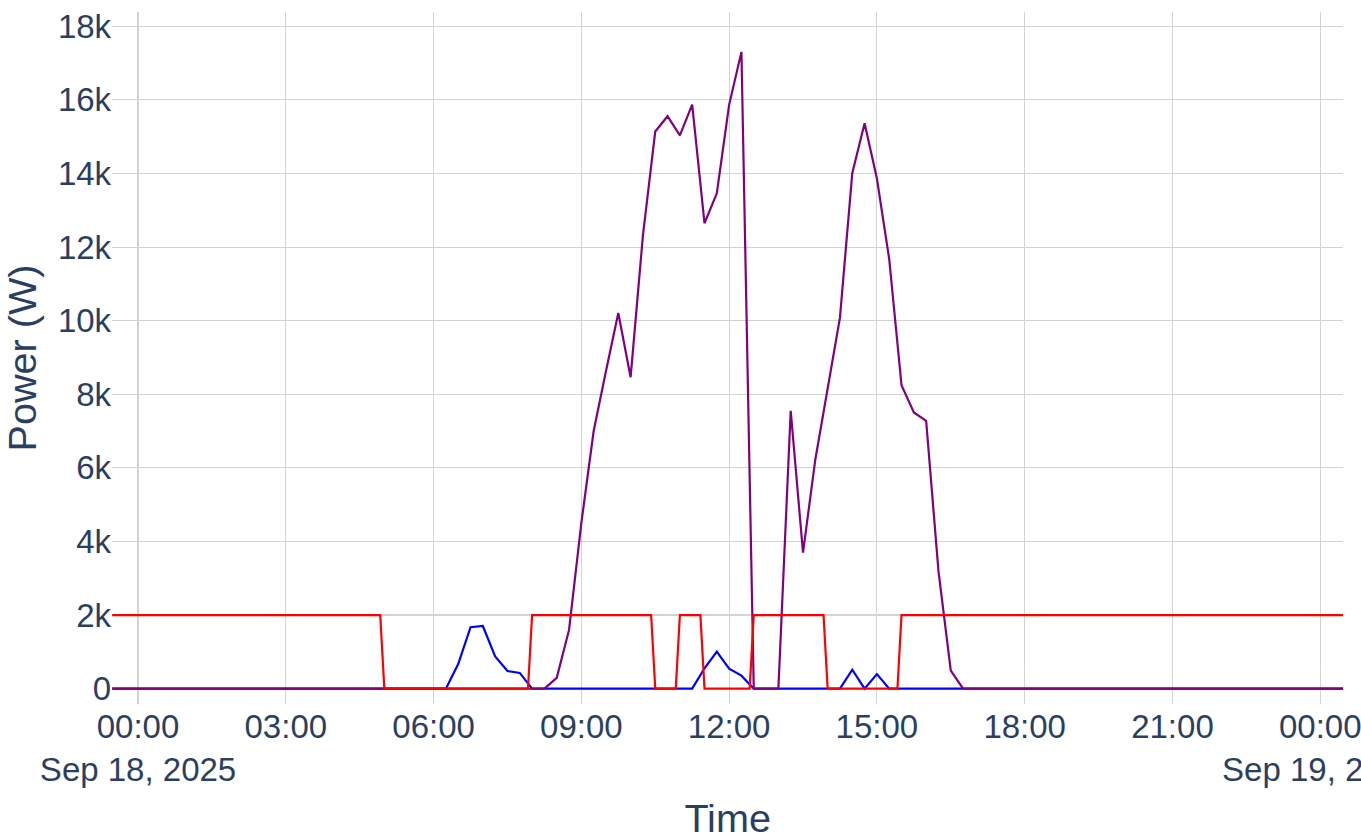}
    \caption{Typical day with good forecast, and successful control of a WSH. The red curve represents the enabling curve (when it is 0, the WSH can load), the blue curve is the power consumption of the WSH, and the purple curve is the surplus curve.}
    \label{fig:control-happy-path}
\end{figure}

\textbf{Fixed Price Household.} The control scheme and operation of the WSH for a household with a fixed energy supplier tariff is illustrated for a typical day in \refig{control-happy-path}. 
~The red curve represents the flexibility orders sent to the WSH. When the red curve equals 0, the WSH is allowed to charge; otherwise, it is not permitted to consume electricity. The blue curve represents the power consumption of the WSH. In \refig{control-happy-path}, it can be observed that the WSH only charges during authorised periods. The purple curve represents the available community PV surplus. As shown, the flexibility orders enable the WSH to charge during periods of surplus, indicating that the forecast was accurate. The time window from 5 AM to 8 AM was defined as a comfort window to ensure adequate hot water availability for household members in the morning. In summary, on the day shown in \refig{control-happy-path}, the control pipeline operated as intended, and the WSH charged using solar energy, except during the comfort window.

\begin{figure}[!h]
  \centering
   \includegraphics[width=1\linewidth]{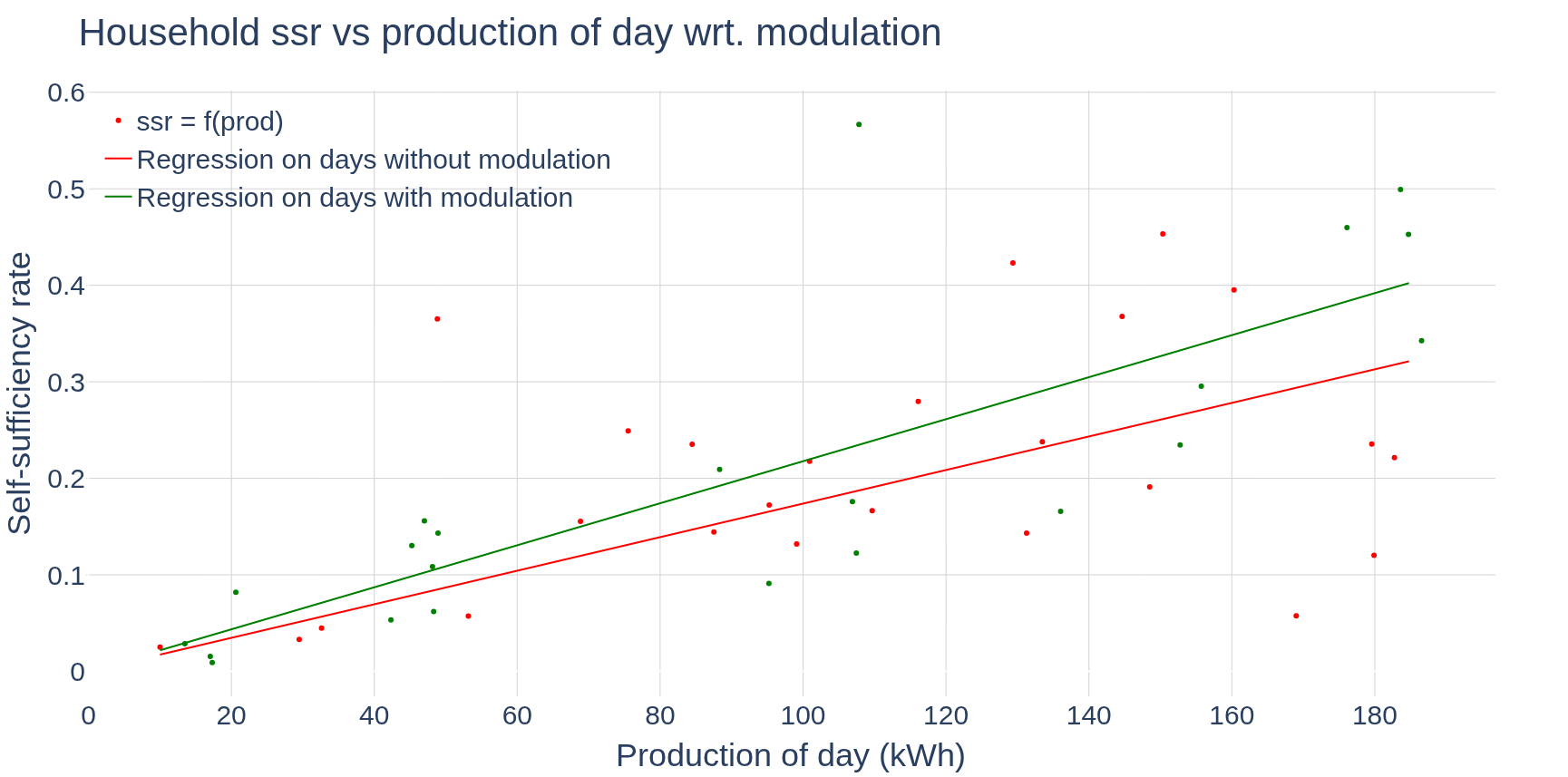}
  \caption{Evolution of self-sufficiency rate for one household on Fixed tariff as a function of solar PV production, with (green) and without (red) flexibility.}
  \label{fig:base_ssr_function_prod_reg}
 \end{figure}
 
We now evaluate the overall efficiency of the system for this WSH. In \refig{base_ssr_function_prod_reg}, we compare the self sufficiency rate (y-axis) as a function of total solar PV production (x-axis) for days with flexibility orders sent (green) and without flexibility orders sent (red). The fitted linear regression curves indicate that self sufficiency increases nearly 50\% faster with PV production on days when flexibility is used compared to days without flexibility activation. Namely, the self-sufficiency rate (as a percentage) gains 10 percentage points on days when orders are sent, every time solar production increases by 50 kWh, while it only gains 5 percentage points on days where no orders are sent, which shows the positive influence of the flexibility scheme.
In terms of economic gains, and making the favorable assumption that all the WSH self-consumption on surplus periods was attributable to the system, the WSH bill reduction amounted to 3.1€/month over the experimentation period (Autumn), computed as the difference between the WSH bill using the community surplus (including network charges and taxes), and what it would have been if it had consumed the same energy directly from the grid, at the Fixed price. This can be extrapolated to an annual gain of 18.6€ by considering that community surplus will only happen during 6 months of the year. However, this assumption does not take into account the fact at least some consumption would have taken place during surplus periods anyway, and is therefore an upper bound on the efficiency of the system.
Results could have been stronger without the inclusion of the comfort window, which led to most hot water production occurring early in the day and consequently reduced the effectiveness of the flexibility system during later periods, and if the experiment had been conducted in summer or over the whole year.

\textbf{ToU Households.} 
Unfortunately, the control of ToU households started late in October due to technical deployment constraints. October corresponds to a period when community surplus was considerably reduced. Furthermore, as previously mentioned, the shift of ToU households’ WSH was performed from one off peak period to another. As a result, ToU households’ WSH were activated only at 12:30, the start of the off peak period, and most units were fully charged by the end of that period, leading to minimal financial gains.
To take advantage of the WSH monitoring that began in September, we estimated the gains that would have been obtained by shifting WSH consumption from 12:30 to 13:30, when community surplus was available, assuming the control experiment had started in September for the ToU WSH. We reconstructed the WSH load profiles that could have been achieved by leveraging community surplus if control had begun in September.
These computations differ from the previous simulations. They are based on the actual monitored WSH consumption recorded by the Voltalis metering equipment, whereas the earlier simulations estimated WSH consumption from total household consumption. We computed an overall reduction in surplus of 15 kWh for the second half of September, with an average of 4.5 active WSH during that period. This corresponds to a monthly saving of 1.1 €/household for September. The value is significantly lower than in the simulation because only the midday WSH peak was shifted (which represents one third of the daily WSH consumption) and the volume of community surplus in September 2025 was limited.

\subsection{Technical challenges in the field deployment of flexibility activation}
From this experiment, we identified the four following issues which flexibility in CSC operations faces.
\textbf{User Engagement.} The first challenge in flexibility-enabled CSC schemes is driving up user engagement. Indeed, we only managed to recruit about 25\% of the members of the operation for this experiment, while members of a CSC operation already tend to be more energy-aware than the general public. We believe this is linked with the complexity entailed, with respect to the gains to be expected. The sole installation of the Voltalis solution for members already represents some \gu{cognitive load} \cite{dellavalle2020nudging}, first due to the need to make an appointment, and being present when the technician comes. Moreover, understanding the logic of the system is not straightforward for non-professional users. Now, the efforts to join a flexibility system must be put in comparison with the gains which they may expect from it. Even in an optimal scenario as computed in Section~\ref{sec:simulation_flex}, members may only gain 70€ a year. This represents around 6€ a month, meaning the gains may be hardly visible for users. Therefore, while there are gains to be made, they may not seem significant enough to justify the required efforts, despite the solution being fully automated once deployed. The activation device however goes beyond the use in water heaters, and also concerns heating, which could lead to higher financial benefits for users. This aspect however was not included in this study, as we focused solely on the specific flexibility of WSHs, however studying the flexibility solution from a holistic point of view may represent an interesting direction of future research.

\textbf{Flexibility Gains with respect to Pipeline Set-Up and Operations.} Moreover, setting-up the flexibility pipelines comes with associated costs. Now, the fact the Voltalis solution is profitable per se means the costs to equip the WSH need not be offset by the flexibility gains. But the gains to be made by targeting local solar surplus still need to be compared to the costs entailed by the software development needed (forecast tool, and specific control logic).

\textbf{Maintaining user comfort.} One issue we had in the optimisation of the water heater consumption was the important consumption in the comfort window, which took place before sunrise (for Fixed tariff consumers) and at 2AM for ToU users and was designed to preserve user comfort (so that people had hot water in the morning). However, the WSH load during this window was significant, reducing subsequent charging needs during surplus periods. As a result, limited solar energy was ultimately consumed during those surplus windows. More granular optimisation strategies, incorporating water temperature dynamics and individual hot water usage patterns, may improve performance in this respect. Further research will explore this direction.

\textbf{Being Conservative or Managing Financial Risks.} Triggering appliance usage shifts risks turning consumptions priced at off-peak tariffs into consumptions priced at peak tariffs, as we explained, due to the risk of forecasts being inaccurate. To mitigate this risk, one would need to control for the probability of making forecast errors, and define a suitable time horizon over which the system would remain profitable on average. Even if this could be done, this would only compound the issues with user engagement which we discussed before. However, to reduce surplus significantly in CSC operations with high percentages of users on ToU tariffs, such developments would be required.

\textbf{Case with a Larger Number of Consumers}
In a scenario with higher flexibility uptake, community surplus may not be sufficient to supply all WSH simultaneously. In the simulation, a pro rata allocation rule was applied, whereby each WSH received a share of the available surplus energy proportional to its demand.
In practice, WSH control is less granular and does not allow perfectly fine tuned modulation of power. The units are controlled using pulse width modulation with a 5 minute switching interval. Consequently, new control strategies must be developed to ensure fair access to community surplus while respecting these technical constraints.

\subsection{Acceptability of Demand-Side Flexibility in Energy Communities}
To assess the acceptability of flexibility, a survey was sent to the participants at the end of the experimentation period. Two participants reported occasionally experiencing a lack of hot water in the evening. This situation may occur when activation ends due to the absence of surplus generation and the system remains inactive until the next off peak period. Even if the root cause does not lie in the flexibility solution itself, the issue may still be attributed to it, particularly because it is perceived as a new element by users. This perception may reduce user engagement and should therefore be taken into account. Rather than relying solely on the night-time peak to ensure hot water availability in the morning, control strategies could also exploit the end of mid-day and evening off-peak periods to meet any remaining daily hot water demand that was not supplied by local energy. Implementing this approach would require forecasting each household’s daily hot water needs, which will be subject to higher variability than community timeseries forecast. %

When asked about other appliances that could be used for flexibility, participants most frequently mentioned space heating and washing machines. Mobilising these additional loads could increase financial benefits and provide stronger incentives to expand solar PV capacity. Finally, one respondent noted that the concept of flexibility and the overall demand side flexibility framework remains largely unclear to the general public. Greater transparency, clear explanations, and active social engagement by professionals and academics are needed to improve acceptance, particularly when financial gains are relatively modest (e.g., an average bill reduction of 70 €/year from flexibility compared to a standard CSC setup, as reported in Section \ref{sec:simulation_flex}).

\section{\uppercase{Conclusions, discussion}}
\label{sec:conclusion}

This work examined WSH driven flexibility within a CSC scheme through both a theoretical simulation and a real world experiment. Theoretical results based on real annual demand and production data indicate that adding WSH flexibility in a CSC energy community could generate average gains of approximately 70 €/year per household in addition to the gains from setting up a CSC energy community. These findings highlight the value of integrating residential flexibility into CSC energy communities and support sustainable business models for rooftop solar PV investment with local energy sharing. 

The real world experiment implemented a predictive control pipeline to operate WSH systems in volunteer households, representing 25\% of the community. Unlike the simulations, the real world experiment faced technical limitations, including the absence of real time monitoring and control. As a result, WSH operation relied on forecasts of community solar surplus, which required the implementation of a safety mechanism to ensure that no participant incurred financial losses.
For households on fixed tariffs, the real world results showed lower gains than the simulated average. This difference is primarily explained by the experimental period, conducted in September, and by the use of a control rule that allowed WSHs to heat at the beginning of the day at full price to avoid any shortage of hot water. For households on time of use tariffs, the safety mechanism to protect from financial loss had a strong impact, as switching from off peak to peak periods was not permitted, which led to almost no financial benefit for households in the considered month.

Future work should ensure that, on average, flexible households are not exposed to financial losses while still capturing surplus generation on a yearly basis. In addition, higher levels of flexibility participation will require the design of new allocation mechanisms to guarantee a fair distribution of community solar surplus within the technical constraints of the current experimental setup, which does not allow granular control.

\section*{\uppercase{Acknowledgements}}

The first author would like to thank Cyril Joly for his help in understanding, and using, the ARPEGE model weather forecast data. All authors would like to thank the volunteers of the experiment for their interest, and for the valuable feedback they provided. This work was parly supported  by EPSRC project HI-ACT (EP/X038823/2).

\bibliographystyle{apalike}
{\small
\bibliography{biblio_enogrid}}

@article{madrigal2026improving,
  title={Improving energy distribution in collective self-consumption via XGBoost-based allocation coefficients prediction},
  author={Madrigal, Sebasti{\'a}n and Gallinad, Ramon and Vicario, Jose L and Morell, Antoni and Vilanova, Ramon},
  journal={Applied Energy},
  volume={409},
  pages={127469},
  year={2026},
  publisher={Elsevier}
}

@inproceedings{allard2024quantifying,
  title={Quantifying, Activating and Rewarding Flexibility in Renewable Energy Communities},
  author={Allard, Julien and Vall{\'e}e, Fran{\c{c}}ois and De Gr{\`e}ve, Zacharie and Stegen, Thomas and Glavic, Mevludin and Corn{\'e}lusse, Bertrand},
  booktitle={2024 IEEE PES Innovative Smart Grid Technologies Europe (ISGT EUROPE)},
  pages={1--5},
  year={2024},
  organization={IEEE}
}

@article{ercoli2025demand,
  title={Demand response for renewable energy communities: Exploring coordination of prosumer-generated PV and flexible aggregated demand in the Italian framework},
  author={Ercoli, Patricia and Mugnini, Alice and Arteconi, Alessia},
  journal={Energy and Buildings},
  pages={115814},
  year={2025},
  publisher={Elsevier}
}

@article{askeland2025smart,
  title={Smart flexibility in energy communities: Scenario-based analysis of distribution grid implications and economic impacts},
  author={Askeland, Magnus and Bjarghov, Sigurd and Rana, Rubi and Morch, Andrei and Taxt, Henning},
  journal={Smart Energy},
  pages={100184},
  year={2025},
  publisher={Elsevier}
}

@article{luzzati2024energy,
  title={Are energy community members more flexible than individual prosumers? Evidence from a serious game},
  author={Luzzati, Tommaso and Mura, Elena and Pellegrini, Luisa and Raugi, Marco and Salvati, Nicola and Schito, Eva and Scipioni, Sara and Testi, Daniele and Zerbino, Pierluigi},
  journal={Journal of Cleaner Production},
  volume={444},
  pages={141114},
  year={2024},
  publisher={Elsevier}
}

@article{di2023flexibility,
  title={Flexibility of grid interactive water heaters: The situation in the US},
  author={Di Silvestre, ML and Sanseverino, E Riva and Telaretti, E and Zizzo, G},
  journal={Renewable and Sustainable Energy Reviews},
  volume={182},
  pages={113425},
  year={2023},
  publisher={Elsevier}
}

@inproceedings{liu2015single,
  title={Single household domestic water heater design and control utilising PV energy: The untapped energy storage solution},
  author={Liu, Aaron Lei and Ledwich, Gerard and Miller, Wendy},
  booktitle={2015 IEEE PES Asia-Pacific Power and Energy Engineering Conference (APPEEC)},
  pages={1--5},
  year={2015},
  organization={IEEE}
}

@article{d2022exploiting,
  title={Exploiting demand-side flexibility: State-of-the-art, open issues and social perspective},
  author={D’Ettorre, F and Banaei, M and Ebrahimy, R and Pourmousavi, S Ali and Blomgren, EMV and Kowalski, J and Bohdanowicz, Z and {\L}opaciuk-Gonczaryk, B and Biele, C and Madsen, H},
  journal={Renewable and Sustainable Energy Reviews},
  volume={165},
  pages={112605},
  year={2022},
  publisher={Elsevier}
}

@article{le2023developing,
  title={Developing energy flexibility in clusters of buildings: A critical analysis of barriers from planning to operation},
  author={Le Dr{\'e}au, J{\'e}r{\^o}me and Lopes, Rui Amaral and O'Connell, Sarah and Finn, Donal and Hu, Maomao and Queiroz, Humberto and Alexander, Dani and Satchwell, Andrew and {\"O}sterreicher, Doris and Polly, Ben and others},
  journal={Energy and Buildings},
  volume={300},
  pages={113608},
  year={2023},
  publisher={Elsevier}
}

@article{tomat2023insights,
  title={Insights into end users’ acceptance and participation in energy flexibility strategies},
  author={Tomat, Valentina and Ramallo-Gonz{\'a}lez, Alfonso P and Skarmeta-G{\'o}mez, Antonio and Georgopoulos, Giannis and Papadopoulos, Panagiotis},
  journal={Buildings},
  volume={13},
  number={2},
  pages={461},
  year={2023},
  publisher={Multidisciplinary Digital Publishing Institute}
}

@misc{couraud2025fairnessenergydistributionmechanisms,
      title={Fairness of Energy Distribution Mechanisms in Collective Self-Consumption Schemes}, 
      author={Benoit Couraud and Valentin Robu and Sonam Norbu and Merlinda Andoni and Yann Rozier and Si Chen and Erwin Franquet and Pierre-Jean Barre and Satria Putra Kanugrahan and Benjamin Berthou and David Flynn},
      year={2025},
      eprint={2508.16819},
      archivePrefix={arXiv},
      primaryClass={cs.CY},
      url={https://arxiv.org/abs/2508.16819}, 
}

@techreport{HPT_Annex46_France_2020,
  title        = {Task 1 Country Report France},
  author       = {{Heat Pump Centre}},
  institution  = {Heat Pumping Technologies},
  type         = {Report},
  number       = {HPT-AN46-02-03},
  year         = {2020},
  url          = {https://heatpumpingtechnologies.org/content/uploads/sites/53/2020/10/hpt-an46-02-03-task-1-counry-report-france.pdf},
  note         = {Final report, IEA Heat Pumping Technologies TCP Annex 46},
}

@misc{codeenergieL315-2,
  title        = {Article L315-2 - Code de l'énergie},
  author       = {{République Française}},
  year         = {2021},
  howpublished = {\url{https://www.legifrance.gouv.fr/codes/article_lc/LEGIARTI000043213495}}
}

@article{dellavalle2020nudging,
title = {Nudging and boosting for equity? Towards a behavioural economics of energy justice},
journal = {Energy Research \& Social Science},
volume = {68},
pages = {101589},
year = {2020},
issn = {2214-6296},
doi = {https://doi.org/10.1016/j.erss.2020.101589},
url = {https://www.sciencedirect.com/science/article/pii/S221462962030164X},
author = {Nives DellaValle and Siddharth Sareen}
}

@article{article,
author = {Lucas, Alexandre and Jansen, Luca and Andreadou, Nikoleta and Kotsakis, Evangelos and Masera, Marcelo},
year = {2019},
month = {07},
pages = {2725},
title = {Load Flexibility Forecast for DR Using Non-Intrusive Load Monitoring in the Residential Sector},
volume = {12},
journal = {Energies},
doi = {10.3390/en12142725}
}

@misc{cre2024pv,
  author = {{Commission de Régulation de l'Énergie}},
  title = {Tarifs et primes relatifs aux installations PV},
  year = {2025},
  note = {Available online at \href{https://www.cre.fr/actualites/toute-lactualite/la-cre-publie-les-nouveaux-tarifs-et-primes-relatifs-aux-installations-photovoltaiques-implantees-sur-batiment-hangar-ou-ombriere-dune-puissance-crete-installee-inferieure-ou-egale-a-500-kw.html}{https://www.cre.fr/actualites/}}
}

@phdthesis{Bejannin_2020, type={Theses}, title={Optimisation du pilotage d’un parc diffus de ballons d’eau chaude pour la fourniture d’offres de flexibilites au reseau electrique}, url={https://pastel.hal.science/tel-02969168}, abstractNote={In France, electric water heaters represent an important source of flexibility for the grid. This thesis stands within the scope of the deployment of innovative telecommunication solutions which aim to quickly and individually address orders to Joule or thermodynamic electric water heaters. Therefore, the approach consists in proposing a model of an electric water heater sufficiently detailed to allow an evaluation of comfort while saving the calculation resources. The model has a low unit calculation time and can be configured easily to represent the French stock in all its diversity using the French territory description databases (INSEE data). In parallel, a consumer behavior model has been developed to simulate annual hot water draws over short time steps. The behavior model, as well as the equipment, is representative of the actual stock in average, but also in their diversity. This stock is adapted to the optimization of thousands of electric water heaters to achieve flexibility objectives. In a second step, an optimization process based on the use of a metaheuristic algorithm by “particulate swarm” is implemented in order to develop strategies for optimizing the control of water heaters and to propose flexibilities to the grid while taking into account the discomfort of users. The realistic control configurations promised by the telecommunication innovations are tested for the flexibility they could provide to grid operators. Finally, the robustness of the obtained control orders with different drop-off scenarios is evaluated. All models and algorithms are integrated into Smart-E, the energy simulation tool for territories at CES.}, number={2020UPSLM010}, school={Université Paris sciences et lettres}, author={Bejannin, Baptiste}, year={2020}, month=feb }

@Article{solar5040048,
AUTHOR = {Matushkin, Dmytro and Zaporozhets, Artur and Babak, Vitalii and Kulyk, Mykhailo and Denysov, Viktor},
TITLE = {Hourly Photovoltaic Power Forecasting Using Exponential Smoothing: A Comparative Study Based on Operational Data},
JOURNAL = {Solar},
VOLUME = {5},
YEAR = {2025},
NUMBER = {4},
ARTICLE-NUMBER = {48},
URL = {https://www.mdpi.com/2673-9941/5/4/48},
ISSN = {2673-9941},
DOI = {10.3390/solar5040048}
}

@article{Demir2025,
  author    = {Demir, Emrah and Gunal, Serkan},
  title     = {Short-term electricity consumption forecasting with deep learning},
  journal   = {The Journal of Supercomputing},
  year      = {2025},
  volume    = {81},
  number    = {10},
  pages     = {1108},
  doi       = {10.1007/s11227-025-07564-5},
  url       = {https://doi.org/10.1007/s11227-025-07564-5},
  issn      = {1573-0484}
}

@article{Breiman2001,
  author    = {Breiman, Leo},
  title     = {Random Forests},
  journal   = {Machine Learning},
  year      = {2001},
  volume    = {45},
  number    = {1},
  pages     = {5--32},
  doi       = {10.1023/A:1010933404324},
  url       = {https://doi.org/10.1023/A:1010933404324},
  issn      = {1573-0565}
}

@misc{elmy_flexibilite_enr_2024,
  author = {{Elmy}},
  title = {Flexibilité {ENR} : comment générer de nouveaux revenus grâce aux marchés de l'électricité},
  year = {2024},
  howpublished = {Online Webinar},
  url = {https://webikeo.fr/webinar/flexibilite-enr-comment-generer-de-nouveaux-revenus-grace-aux-marches-de-lelectricite},
  note = {Accessed 23 February 2026},
  organization = {Elmy},
  language = {Français}
}

@misc{octopus_ballons_2026,
  author = {{Octopus Energy France}},
  title = {À fond les ballons},
  year = {2026},
  howpublished = {Web page},
  url = {https://octopusenergy.fr/ballons},
  note = {Accessed 23 février 2026},
  organization = {Octopus Energy},
  language = {Français}
}

\end{document}